\documentstyle[preprint,pre,aps]{revtex}

\tighten

\begin{document}

\draft

\preprint{LA-UR-93-4267\ {\bf /}\
MA/UC3M/10/93}

\title{Kink stability, propagation, and length scale competition in
the periodically modulated sine-Gordon equation}

\author{Angel S\'anchez}

\address{Theoretical Division and Center for Nonlinear Studies,
Los Alamos National Laboratory,\\
Los Alamos, New Mexico 87545,
and\\ Escuela Polit\'{e}cnica Superior, Universidad Carlos III de
Madrid,
E-28913 Legan\'{e}s, Madrid, Spain}

\author{A.\ R.\ Bishop}

\address{Theoretical Division and Center for Nonlinear Studies,
Los Alamos National Laboratory,\\
Los Alamos, New Mexico 87545}

\author{Francisco Dom\'\i nguez-Adame}

\address{Departamento de F\'\i sica de Materiales, Facultad de
F\'\i sicas, Universidad Complutense,\\ E-28040 Madrid, Spain}

\date{\today}

\maketitle

\begin{abstract}
We have examined the dynamical behavior of the kink solutions of the
one-dimensional
sine-Gordon equation in the presence of a spatially periodic
parametric perturbation.
Our study clarifies and extends the currently available knowledge on this
and related nonlinear
problems in four directions. First, we present the results of a numerical
simulation program which are not compatible with the existence of a
radiative threshold, predicted by earlier calculations.
Second, we carry out a perturbative calculation which helps interpret
those previous predictions, enabling us to understand in depth our
numerical results. Third, we apply the collective coordinate formalism
to this system
and demonstrate numerically that it accurately reproduces the observed
kink dynamics. Fourth, we report on a novel occurrence of length
scale competition in this system and show how it can be understood by means
of linear stability analysis. Finally, we conclude by summarizing the
general physical framework that arises from our study.
\end{abstract}



\narrowtext

\section{Introduction.}

Technological progress has made possible the fabrication of highly ordered
materials and structures for a very large number of applications. In
parallel to those advances, it has also been realized that the
special properties required for many purposes necessitate
inhomogeneous systems. Here inhomogeneity may mean spatial modulations,
quasiperiodicity, or disorder of several kinds. In addition,
there are other situations in which inhomogeneity is undesirable
but unavoidable. In either case, the study of disordered systems
acquires fundamental importance. This is even more so when the physical
system in
which disorder or inhomogenity is to be studied is
described by a nonlinear model.
Whereas the r\^ole of disorder in linear problems is at least partially
understood, much less is known about nonlinear disordered systems. In fact,
even from a mathematical viewpoint, the understanding of these
models, often related to stochastic partial differential equations
(PDE),
is very limited. Consequently, a great deal of research has
been devoted to this topic \cite{Proceedings,ReviewYura,Reviews}.

A major part of the work done so far regarding nonlinear disordered
systems has been concerned with some particular examples that are amenable
to analytical treatment while capturing some essential physics.
The sine-Gordon (sG) (actually, the whole family of nonlinear
Klein-Gordon equations, including, e.g., the $\phi^4$, double- and
quadratic- sine-Gordon equations) and nonlinear Schr\"odinger (NLS)
equations are often chosen as
very suitable ``canonical'' examples. This is due to
the fact that the basic mathematical structure underlying them is well
known and therefore provides a good starting point for theoretical work.
This reason would not be sufficient if these models were not
also related to
a large number of phenomena that occur in quasi-one-dimensional physical
systems, as is in fact the case. In the context of these two models,
disorder is introduced through suitably chosen perturbation terms (see
\cite{ReviewYura} for an extensive list of physically relevant
perturbations). This is the usual procedure by which inhomogeneity
of any kind is studied: The equation describing the problem is
established,
the terms relevant to the considered physical situation are
identified, and a perturbation to those terms is introduced, representing
the desired kind of disorder. Our viewpoint in this work is more generic:
Although the system we deal with is indeed related to a number of
applications, our aim is that we will be able to gain insight into
underlying
mechanisms of the phenomenology of nonlinear disordered systems.
Therefore, we introduce a simple periodic perturbation which will
allow us to study very interesting and general phenomena, such as
length scale competition, and will provide information relevant to
the more complicated processes occurring in random media (the periodic
potential can be interpreted as a ``color'' of a general noisy one).
The
knowledge obtained
will also be useful to tackle other problems where
detailed studies including analytical treatment are not possible.

In this paper we study the behavior of one-dimensional (1D)
sG kinks when perturbed {\em parametrically}
by a spatially periodic potential. Initially, this was motivated
by our related research (from the above general point
of view) on the sG \cite{sG,sGcc} and NLS \cite{NLS,NLScc} models. As a
preliminary
step to the investigation of sG breather dynamics \cite{sG} on these kind of
potentials, it is natural to first seek a good understanding
of kink dynamics.
Therefore we undertook that study, both analytically and numerically.
Our point of departure was early theoretical work%
\cite{raro,Boris1,Boris2}
on this problem, which we summarize for completeness in Sec.\ II.
In particular, it
had been predicted that a certain critical velocity exists at which
the radiative power emitted by the kink would diverge. Below that
critical velocity, radiation would be zero, and above it, it would
decrease with increasing speed (see Ref.\ \onlinecite{ReviewYura} for a
summary). Those results were obtained at a time where the main aim
was to develop a perturbative approach to deal with soliton problems.
That, and the fact that computers were not the easily available tool they
are nowadays, meant that those results were never analyzed in depth or
numerically checked. Therefore, as a first stage of our study,
we devised a number of numerical experiments to check them, and we found
no numerical evidence for the predicted divergence. In view of this
result, we carried out an improved perturbative calculation,
in the sense that it allowed us
to interpret correctly the earlier results in Refs.\
\onlinecite{raro,Boris1,Boris2} and to show that,
although the earlier analyses were correct, the predicted
divergence was actually unphysical. This theoretical analysis
is reported in detail
in Sec.\ III: A preliminary short report has been given elsewhere
\cite{otro}. The work done to that point suggested to us the idea that,
opposite to what was believed to date, sG kink dynamics on a periodic
potential could be essentially that of a (pseudo-relativistic)
particle. We thus applied a simple
collective coordinate formalism to the problem, and it turned
out to describe soliton behavior very accurately, even predicting
unexpected new phenomena. The analytical approach
and the numerical simulations are contained in Sec.\ IV.
Finally, to complete our program studying length scale competition
and its effects in nonlinear disordered systems, we performed further
numerical
experiments to clarify whether robust objects like kinks, which
according to our collective coordinate theory behave mostly like particles,
can still exhibit the destabilizing effect of length scale competition.
We found that this was
actually the case. Furthermore, the simplicity of kinks allowed us to
carry out a (numerical) linear stability analysis which provided us
with a clear explanation for the numerically observed features. We
collect our results on this question in Sec.\ V. To conclude, we summarize
the facts that we have learned, which considerably enhance the understanding
of sine-Gordon kink propagation in disordered media and shed light in the
so far unexplained phenomenon of length scale competition.
Our results are also of relevance to
many nonlinear systems of physical interest, mainly in three directions:
First, all non-numerically validated or
non-physically interpreted predictions obtained through perturbative
calculations should be treated with a degree of caution. Second, the
collective coordinate formalism yields a very simple and accurate way
to deal with perturbed nonlinear problems, especially those in
which the perturbation enters parametrically rather than additively.
And third, length scale competition is an ubiquitous phenomenon that may be
responsible for many instabilities arising in
different nonlinear disordered systems.

\section{Brief summary of previous results.}

We start by describing the picture of sG kink propagation
on parametric
periodic potentials that has been accepted to date. The problem,
that has been studied by Mkrtchyan and Shmidt
\cite{raro} and Malomed and Tribelsky \cite{Boris1,Boris2} is given
by a perturbed sG equation of the form
\begin{equation}
\label{PsG}
u_{tt}-u_{xx}+[1+\epsilon\cos(kx)]\sin u = 0,
\end{equation}
(modeling for instance
a long Josephson junction with modulated critical current, to mention
just one application); the
question posed was whether kinks can propagate freely
in such a system, and
if so, to describe this propagation. We will only
record here a short summary of previous work. The reader is
referred to the original papers \cite{raro,Boris1,Boris2} for details.

Mkrtchyan and Shmidt \cite{raro} used a Green-function
perturbation technique (GFPT). They derived a linearized equation for
the first order correction to a kink moving with constant velocity,
computed the Green function corresponding to its homogeneous version,
and then used it to obtain the desired correction by integrating the
source term with that Green function. They then noticed that radiation
appeared only above a critical kink velocity
$v_{\scriptstyle\rm thr}=(1+k^2)^{-1/2}$. At that particular value, the
correction diverges, and the authors explain that their calculation
become invalid in that region, as of course, it assumed the correction
was small.
On the other hand, the approach of Malomed and Tribelsky
\cite{Boris1,Boris2} was quite different. Its basis was
the Inverse Scattering Perturbation Theory (ISPT). A meaningful
summary of this kind of calculation would be quite lengthy, and
hence we will omit it here, referring the reader to Ref.\
\cite{ReviewYura} which is mostly devoted to describing ISPT in
detail. Let us just mention that the idea is that, if the amount
of radiation emitted by the kink is small, as it should be if the
perturbation is small, then the spectral density of the emitted
energy can be computed following a Taylor expansion, and the total
radiated energy is then derived by integration over all modes.
Again, the result was that
there was a critical velocity $v_{\scriptstyle\rm thr}=(1+k^2)^{-1/2}$
such that kinks traveling with velocities $v<v_{\scriptstyle\rm thr}$ did
not emit any radiation at all, whereas in the opposite case the amount
of emitted radiation decreased as $v\to 1$, diverging when
$v=v_{\scriptstyle\rm thr}$. Always within
the framework of ISPT,  Malomed and Tribelsky
\cite{Boris2} were also able to show that dissipation could play a
regularizing r\^ole, suppressing the divergence.
As the results in Ref.\ \cite{raro} agreed with those in Refs.\
\cite{Boris1,Boris2},
the existence of this threshold for radiation with its associated
divergence was accepted, and the question of kink propagation on
periodic potentials was regarded as basically solved. As mentioned
above, it has to be
borne in mind that the main issue of those early researches
was to establish
the proper foundations for a perturbative theory for solitons.
Hence, the question of the physical meaning and origin of
the divergence was not addressed. Another unexplained point arises
already from ISPT, which allows computation of the radiation
nature. When this is done in our case, the radiation wavenumbers
turn out to be related to the perturbation one by a complicated equation
(see, e.g., \cite{ReviewYura}),
which, in particular, implies that radiation is emitted with
a non-intuitive wavenumber $k^{-1}$ at the divergence.
This prediction is difficult to understand physically. Let us recall
at this point that a particlelike picture of kink propagation had
been developed and had been largely successful so far \cite{viejos}
when compared to numerical experiments. If ISPT predictions for
the radiation wavenumbers were true, the
reason for them must come from the wave nature of kinks. Consequently,
the particle picture should be regarded as a major simplification
and valid only in limited cases.

\section{Kink propagation on periodic media.}

\subsection{Numerical results.}

With the above scenario (and the question it poses) in mind, we carried
out a number of numerical simulations looking in the first place for
the proposed
threshold.
All the simulations we will be reporting on throughout the paper
have been carried out taking
periodic boundary conditions. The integration was performed with two
different procedures, an adaptation of the energy conserving
Strauss-V\'azquez finite-difference scheme \cite{yonum}
and a fifth order, adaptive stepsize,
Runge-Kutta integration \cite{Recipes} of the discretization of the PDE.
The results were independent of the procedure, which is a satisfactory
checking.
We performed a careful search, paying attention to the
fact that the predicted value was a first order calculation, and that
it may not be quantitatively
accurate. On the other hand, the finite width of the simulated system
may also be of relevance at this point, as its radiation spectrum
structure is not identical to the continuum, infinite system (in
particular, the lowest frequency in the model is restricted to be
$\omega_{min}^2=1+(2\pi/L)^2$, $L$ being the length of the system).
Hence, we monitored the amount of radiation
emitted by the kink
by making simulations with many different initial conditions, sweeping
a range of initial velocities;
if there was a threshold somewhere, there should
be a change in the radiating power of the kink as it moved through it.
The result was
negative: No evidence for a threshold was found, even when the search
was performed for a large range of initial velocities with a resolution of
$10^{-2}$ for some choices of $k$. Examples of the outcome of the
simulations are shown in Fig.\ \ref{nodiv} for three values of the
potential wavelength, of the order (a) and smaller (b, c) than
the kink width ($\sim 6$ in our dimensionless units) at
$v=v_{\scriptstyle\rm thr}$.
It has to be stressed that
the predicted divergence does not depend on the strength of the
perturbing potential, $\epsilon$, but we also tried to make the effect
more
visible by increasing this parameter. Indeed, in Fig.\ \ref{nodiv},
$\epsilon=0.4$, a value that is not very small, and the kinks seem
unaffected except for a small amount of radiation and an oscillatory
motion superimposed on its trajectory, which is
shown in Fig.\ \ref{centers}.
It is interesting to note that the
kink traveling on the short wavelength potential (c) appears not to
be affected at all. This will be understood by means of the collective
coordinate approach in Sec.\ IV.
On increasing $\epsilon$ further, trapping behavior takes
place, i.e., kinks are trapped by the potential and cannot propagate,
but there is no strong emission of radiation (for an example, see Fig.\
\ref{coll}(b), which will be discussed later). Actually, this trapping
can be of two very different kinds, as we will discuss in sections IV
and V. Another interesting remark is that we also observed that kinks
always emit radiation, even when moving at a very low velocity, far below
the predicted threshold. A similar result arises from the work of Peyrard and
Kruskal on highly discrete sG systems \cite{Michel}, where kinks propagate
on the periodic potential coming from the Peierls-Nabarro barrier,
although this comparison should not be taken too literally as there
are some differences between both problems, like the existence of a
maximum allowed frequency in the discrete one, for instance.
It thus becomes evident that the features of
kink propagation on periodic potentials are qualitatively different from
the above perturbative analytical results. Interestingly, numerical
simulations on a similar perturbation of the $\phi^4$ problem
\cite{Madrid} seem to confirm the absence of this divergence. We will
elaborate more on this when presenting our conclusions in
Sec.\ VI.

\subsection{Theory.}

In order to gain insight into the numerical observations, we developed a
new perturbative approach for this problem, following a similar approach
to that given by Fogel {\em et al.\/}\cite{viejos}. To this end, we
perform a Lorentz transformation and rewrite (\ref{PsG}) in the rest
frame of the soliton (i.e., the reference frame moving with the speed of
the unperturbed soliton, $v$)
\begin{equation}
\label{PsGLorentz}
u_{tt}-u_{xx}+\{1+\epsilon\cos[k\gamma (x+vt)]\}\sin u = 0,
\end{equation}
with $\gamma\equiv (1-v^2)^{-1/2}$ the Lorentz factor.  Here we consider
$\epsilon \ll 1$, so the perturbative term may be treated by assuming a
solution of the form
\begin{equation}
u(x,t) = u_{v}(x) + \epsilon u^{(1)}(x,t),
\label{firstorder}
\end{equation}
where $u_{v}(x)\equiv4\tan^{-1}(e^x)$ is the unperturbed sG kink.
For completeness, we now recall how the most appropriate basis
in which to expand $\epsilon u^{(1)}(x,t)$
is obtained. Introducing the {\em Ansatz} (\ref{firstorder})
in Eq.\ (\ref{PsGLorentz}) without the perturbation term,
linearizing in the small quantity $u^{(1)}(x,t)$, and separating time
and space by introducing $u^{(1)}(x,t)=f(x)e^{-i\omega t}$, we
are left with the following eigenvalue problem for $f(x)$:
\begin{equation}
\label{extra1}
\left[-{d^2\phantom{\mbox{$x^2$}}\over dx^2}+ \left(1 - 2\mbox{\rm sech}^2
x \right)\right] f(x) = \omega^2 f(x).
\end{equation}
This is a well known eigenvalue problem \cite{extras}; there exists exactly
one bound state, with $\omega_b=0$, and a continuum of scattering states
with $\omega_{\kappa}^2 = 1+\kappa^2$; the corresponding
normalized eigenfunctions are
\begin{mathletters}
\label{extra2}
\begin{eqnarray}
f_b(x) & = & 2\>\mbox{\rm sech}\, x \\
f(\kappa,x) & = & {1\over\omega_{\kappa}\sqrt{2\pi}}
e^{i\kappa x}\left( \kappa +
i\tanh x\right).
\end{eqnarray}
\end{mathletters}
These eigenfunctions have a very clear physical meaning. The bound
state $f_b(x)$ is associated to the Goldstone
translation mode of the soliton, whereas the
continuum eigenfunctions $f(\kappa,x)$ are the radiation modes
(see \cite{viejos} for a detailed discussion).
Besides, these functions form an orthogonal basis,
 since the corresponding operator is
self-adjoint. We will make use of this fact to deal with our problem.
In terms of this basis,
the first order correction can be split into two parts, namely,
\begin{equation}
u^{(1)}(x,t)=u^{({\scriptstyle\rm trans})}(x,t)+
u^{({\scriptstyle\rm rad})}(x,t),
\label{split}
\end{equation}
where
\begin{mathletters}
\label{u1}
\begin{eqnarray}
u^{({\scriptstyle\rm trans})}(x,t)&=&{1\over 8}\phi_b(t)f_b(x),
                  \label{u1a} \\
u^{({\scriptstyle\rm rad})}(x,t)&=&\int^{\infty}_{-\infty}d\kappa
\>\phi(\kappa,t)f(\kappa,x). \label{u1b}
\end{eqnarray}
\end{mathletters}
To find the amplitudes $\phi_b(t)$
and $\phi(\kappa,t)$, one again introduces the {\em Ansatz}
(\ref{firstorder}) in Eq.\ (\ref{PsGLorentz}), linearizes and
Fourier transforms in time; subsequent projection yields
\begin{mathletters}
\label{phis}
\begin{eqnarray}
\label{phib}
\ddot{\phi}_b(t) & = & 4\int^{\infty}_{-\infty}dx\>\cos[k\gamma (x+vt)]
{\sinh x\over\cosh^3x}, \\
\label{phikappa}
\ddot{\phi}(\kappa,t) + (1+\kappa^2)\phi(\kappa,t) & = &
2\int^{\infty}_{-\infty}dx\>\cos[k\gamma(x+vt)] {e^{-i\kappa x} (\kappa
-i\tanh x)\over \sqrt{2\pi(1+\kappa^2)} } {\sinh x\over \cosh^2 x}.
\end{eqnarray}
\end{mathletters}

It now remains to solve Eqs.\ (\ref{phis})
 and invert the various transforms
needed to arrive at them. In the following,
we discuss translation and radiative parts in (\ref{split})
separately. Let us start with the simplest one, i.e., the translation
mode contribution. Note that (\ref{phib}) is, after performing the
integration, nothing but the Newton's law for a time-dependent force.
Its solution may be readily found, and finally one obtains
\begin{equation}
\label{trans}
u^{({\scriptstyle\rm trans})}(x,t)=
{\pi\over 2v^2\sinh(k\gamma\pi/2)}\sin(k\gamma vt)\> \mbox{\rm sech}\, x.
\end{equation}
Recalling that we are working in the unperturbed soliton reference
frame, this is a localized oscillatory motion superimposed on its
otherwise constant trajectory.  Now, let us remark that the prefactor
implies that short wavelength ($k\to\infty$) perturbations will have no
effect on the motion of the center of the soliton, which is also in good
agreement with our simulations in Fig.\ \ref{centers}.  This behavior
can be understood in terms of a ``smoothing'' of the potential: The
kink, having a width much larger than the perturbation wavelength,
experiences only an effective averaged force, whose amplitude vanishes
exponentially for large $k$ (see Sec.\ IV; see also
related comments in \cite{sG,NLS}).

Equation (\ref{phikappa}) for the $\kappa$-mode radiative contribution
can also be solved.  After computing the integral in the right hand side
of Eq.\ (\ref{phikappa}), one is left with the Newton's law for a
forced harmonic oscillator.  This allows the determination of
$\phi(\kappa,t)$, and substitution of it in Eq.\ (\ref{u1b}) to find
the total radiative contribution:
\begin{eqnarray}
\label{totrad}
u^{({\scriptstyle\rm rad})}(x,t)\equiv
{1\over 4}\> \left[\tanh x - {\partial
\phantom{x}\over\partial x}\right] & \times & \\
\times \int_{-\infty}^{\infty}d\kappa
{1+\kappa^2-k^2\gamma^2\over(1+\kappa^2)(1+\kappa^2-k^2\gamma^2v^2)}
& & \left[ {e^{ik\gamma v t}\over \cosh[\pi(k\gamma -\kappa)/2]}
+{e^{-ik\gamma v t}\over \cosh[\pi(k\gamma +\kappa)/2]}\right]
e^{i\kappa x}.\nonumber
\end{eqnarray}
It is possible to deal with the integral in (\ref{totrad}) in the complex
plane: When $x>0$, in the upper half plane, and when $x<0$ in the
lower half plane. The pole structure of the integrand will
completely determine
the total radiative contribution. In particular, we will see
that radiation only appears for some special values of the system
parameters.

We take $x>0$ in what follows (the opposite case is treated in the same
way).  Accordingly, the integral has to be analyzed in the upper half
complex plane.  The pole structure of the integrand is depicted in Fig.\
\ref{poles}.  All poles are simple, and their locations are $z_0\equiv
+i$, $z_1\equiv i\alpha\equiv +i\sqrt{1-k^2\gamma^2v^2}$, and
$z_n^{\pm}\equiv \pm\kappa\gamma + i (2n+1)$, $n$ being a non-negative
integer. For the sake of clarity we treat each pole separately.

i.  The first pole, $z_0=+i$, is constant, and does not change when the
system parameters change.  Since this pole is purely imaginary, it is
immediately seen that the contribution of the residue at $z_0$ is
exponentially localized around the kink center.  This term does not give
rise to any radiation, but rather to time-dependent corrections of the
kink shape.

ii.  The family of poles $z_n^{\pm}$ depends on the perturbation
wavenumber $k$ and on the kink velocity through the Lorentz factor
$\gamma$.  However, they always have a positive imaginary part, thus
leading again to exponentially localized contributions.  Therefore, the
$z_n^{\pm}$ poles also do not produce any true radiative correction.

iii.  The remaining pole is the key one.  If $\alpha^2\equiv 1-
k^2\gamma^2v^2>0$, the same reasoning applied to the other poles holds,
and there is no radiation.  It is worth mentioning that localized
oscillations around the kink center, predicted from the contributions of
$z_0$, $z_1$ ($\alpha$ real), and $z_n^{\pm}$, were already evident in our
numerical simulations, as shown in Fig.\ \ref{nodiv}.  For fixed $k$,
as $v$ increases, the pole moves down the imaginary axis, and at
the critical value $v_{\scriptstyle\rm thr}\equiv (1+k^2)^{-1/2}$ it
lies at the origin of the complex plane.  For kink velocities
$v>v_{\scriptstyle\rm thr}$ the pole is purely real, and then it does
give rise to a radiative contribution, whose form is given by (with
$\beta\equiv \sqrt{k^2\gamma^2v^2-1}$ a real number)
\begin{equation}
\label{yoquese}
u^{({\scriptstyle\rm rad})}_{\beta}={\pi\over 4\gamma^2 v^2}\left(
1-{i\over \beta} \tanh x\right)
\left[ {e^{i(k\gamma v t+\beta x)}\over \cosh[\pi(k\gamma -\beta)/2]}
+{e^{-i(k\gamma v t-\beta x)}\over \cosh[\pi(k\gamma +\beta)/2]}\right].
\end{equation}
This expression tells us that radiation occurs whenever $\beta$ is real
($v>v_{\scriptstyle\rm thr}$), and this radiation is the
superposition of two linear waves of different amplitudes, travelling in
opposite directions but with the same phase velocity.

\subsection{Discussion.}

To this point, it appears that our perturbative calculation
leads exactly to the same prediction
as those in \cite{raro,Boris2}, namely that there is a critical velocity
$v_{\scriptstyle\rm thr}\equiv (1+k^2)^{-1/2}$ below which kinks do
not radiate and above which they do. At that precise velocity,
the amplitude of the emitted radiation diverges; notice that $\beta$
vanishes as $v$ approaches
$v_{\scriptstyle\rm thr}$ from above and consequently the
prefactor in Eq.\ (\ref{yoquese}) goes to infinity.
However, this apparent equivalence is not so. The crucial difference
arises when one
looks more carefully at Eq.\ (\ref{yoquese}): As
$v_{\scriptstyle\rm thr}$ is approached, not only the amplitude of
the emitted wave diverges, but also its wavelength $2\pi/\beta$.
Then, we are faced with something
similar to an ``infrared'' divergence, and usually those do not
have a real physical meaning. We will show immediately that this is
indeed the case here, but let us first comment on the reasons why
our calculation provides us with this physically relevant result
that was not transparent in the previous approaches. As to the GFPT
computation \cite{raro}, they compute the first order
correction to the field much as we do here (actually the two approaches
are basically the same in the beginning), but they do not use
the natural translation mode-radiation basis, so they can not separate
the different contributions and are therefore led to an expression
they can not analyze in detail; as we already pointed out,
they merely remark that their calculations
are invalid in the vicinity of the divergence, as they assumed the
correction should be small. On the other hand, ISPT \cite{Boris1,Boris2}
yields a different result than ours in spite of using a suitable
basis because the integration over $\kappa$ is made in an incoherent
fashion, i.e., integrating over emitted {\em energy} instead of emitted
{\em amplitude} (we notice in passing that many ISPT results are obtained
by this same means). When the integration over radiation modes is
made coherently as shown here, the result changes due to the superposition
of different
modes. These reasons lead us to believe that, although admittedly
the early perturbative work
was mathematically sound, the calculation
we present here is the {\em physically} correct first order result.

Now that we have a reliable perturbative calculation, we need
to understand what is the nature of the divergence.
 To make progress, it is very important to
turn to the form of our starting Eq.\ (\ref{PsG}) {\em with dimensions},
namely
\begin{equation}
\label{PsG2}
u_{tt}-c_0^2 u_{xx}+\omega^2_0 [1+\epsilon\cos(kx)]\sin u = 0,
\end{equation}
where $c_0$ and $\omega_0$ are a velocity and a frequency characteristic
of the particular physical context. Redoing the calculations with
dimensions transforms the divergence condition $k\gamma
v_{\scriptstyle\rm thr} = 1$ into $k\gamma_0 v_{\scriptstyle\rm thr} =
\omega_0$ [$\gamma_0=(1-v^2/c_0^2)^{-1/2}$]. This immediately
clarifies what happens: The divergence occurs when the velocity of the
kink is such that the time it takes to travel through a wavelength of the
potential, $T_0=\lambda/(\gamma v_{\scriptstyle\rm thr})$, $\lambda =
2\pi/k$, is exactly the period of the lowest frequency
phonon, $T_0=2\pi/\omega_0$.
If the velocity is lower than $v_{\scriptstyle\rm thr}$, the kink will
not be able to excite phonons, whereas when its velocity is higher it
can and will subsequently radiate. At $v_{\scriptstyle\rm thr}$, the
excited radiation is that of the lowest phonon, and it has infinite
wavelength and velocity, as predicted by our calculation. This
natural picture of kinks exciting radiation according to the frequency
of their propagation through a potential wavelength becomes therefore
the likely candidate to explain the divergence. On the other hand, now
it also becomes clear the divergence of the energy at $v_{\scriptstyle
\rm thr}$: It diverges because of the infinite contribution arising
from the infinite wavelength mode when integrated over the whole
$x$ axis. This agrees with GFPT and ISPT results whose only difficulty
was not to specifically identify the mode responsible for the divergence.

In spite of this clarification, the most significant question is
not answered yet: Why numerical simulations do not agree with this
calculation, which seems to allow for a simple and physically reasonable
interpretation? By looking again at Figs.\ \ref{nodiv} and
\ref{centers}, it is
easy to realize that the flaw of the perturbative calculation is at
its very root: We are computing first order corrections around a kink
moving at a {\em constant velocity} $v$, and this condition never
holds. Whatever the starting position of the kink is, it will
behave like a particle in the sense that it will be accelerated or
decelerated depending on whether it travels towards a minimum or a
maximum of the potential. In fact, the translation mode correction
itself is describing this: The kink velocity, in its reference frame,
is not zero but rather it oscillates between positive and negative
values. It is not a surprise, then, that first the resonance condition
we have obtained is never matched, and second that the kink emits
radiation at any velocity, because it is accelerating or
decelerating.
Of course, we should note that this is a perturbative
calculation including only first order terms; the possibility still
remains that the divergence is suppressed by higher order nonlinearities.

\section{Collective coordinate approach.}

The above numerical results and the subsequent perturbative calculation
strongly suggest that sG kinks behave as point-like particles in the
presence of a periodic
parametric potential like the one we deal with here. Therefore,
it is natural to try to describe those results by means of the collective
coordinate formalism. This approach was first proposed in \cite{viejos},
and it has been applied recently to sG breathers on periodic
potentials \cite{sGcc}
as well as to NLS \cite{NLScc} equations with the same perturbation. In
both cases the analytical predictions turned out to be in very good
quantitative agreement with numerical simulations: For instance, in
Ref.\ \cite{sGcc} the threshold for breather breakup into a kink-antikink
pair was predicted with an accuracy better than 0.1\%. On the other hand,
the calculation in Ref.\ \cite{NLScc} predicted the appearance of the
so called ``soliton chaos,'' verified by simulations of the full PDE.
In our present
problem, the advantage we have is that, due to the simpler nature of the
kink, we will be able to compute the effective potential not only for
kinks at rest but also for moving kinks.

The basic idea of the collective coordinate formalism is very simple:
To reduce a complicated problem with an infinite number of degrees of
freedom, posed in terms of a PDE, to a much
less complex problem with a few degrees of freedom (and correspondingly
described in terms of ordinary differential equations, ODE's). There are a
number of ways to do this, and different quantities can be chosen to
play the r\^ole of collective coordinates describing the motion of the
nonlinear excitation as a whole. For our problem, it is enough to simply
consider the kink center as our collective coordinate for the kink.
Its motion will be then governed by an effective potential that can be
computed by integrating the perturbative contribution to the hamiltonian
over the kink profile, i.e.,
\begin{equation}
\label{intveff}
V_{\scriptstyle\rm eff}(x_0,t) = \epsilon \int_{-\infty}^{\infty} dx
\left[1-\cos u_v(x-x_0,t)\right]\cos kx,
\end{equation}
where $u_v(x-x_0,t)$ denotes now a kink moving with constant velocity
$v$ and centered at $x_0$. This integral can be easily evaluated and
yields
\begin{equation}
\label{veff}
V_{\scriptstyle\rm eff}(x_0,t) = 2\epsilon{k\pi\over \gamma^2
\sinh(k\pi/2\gamma)}\cos[k(x_0+vt)].
\end{equation}
{}From equation (\ref{veff}) we see that the potential experienced by the
particle equivalent to the kink is basically the same perturbation potential
that appears in the PDE (\ref{PsG}), although the
prefactor in front of it is quite complicated. The simplest dependence of
this prefactor is on the wavenumber. It can be immediately seen
that when $k\to 0$ (long wavelength limit) the effective potential prefactor
reduces to $4/\gamma$ and subsequently $V_{\scriptstyle\rm eff}$ becomes
closer to the perturbative one; in the opposite limit, $k\to\infty$,
the $\sinh$ term makes the effective potential vanish exponentially.
This is in
agreement with what we have learned so far: Looking at Fig.\ \ref{nodiv},
it can be seen that the short wavelength potential has no effect on the
kink (c), whereas the motion on long wavelength perturbations resemble
that of a particle on the bare potential. To phrase in the terminology
introduced in Ref.\ \cite{sG}, the behavior of the kink in these cases is
that of a ``bare'' (long wavelength) or a ``renormalized'' (short
wavelength) particle. It is also important
to notice that this result agrees with the perturbative calculation
we described in Sec.\ IIIb [see Eq.\ (\ref{trans})] as it was to be
expected.
There we showed that the correction to
the center of mass motion was basically an oscillatory term, implying
that the velocity of the center of mass oscillates around some mean
value. This is precisely the same kind of trajectory followed
by a point-like particle in the potential in Eq.\ (\ref{veff}) (at
least if the velocity is not too close to 1).

Nevertheless, it is worth pursuing this agreement a bit further, by
studying the threshold for kinks to propagate in this kind of potential.
The easiest way to compute the threshold is by equating the kinetic
energy of the kink to the maximum of the effective potential, provided
we restrict ourselves to the non-relativistic limit ($v^2$ not too
close to 1) to keep the kink mass constant. This will
give us the maximum potential height over which a kink that starts
from a point at which the perturbation is zero with a certain
velocity is able to overcome the
nearest top
point. Using the fact that
the mass of a not too fast kink is 8 in our units, we find that
the threshold is given by
\begin{equation}
\label{thre}
\epsilon_{\scriptstyle\rm thr}
 = {2 v^2\gamma^2\over k\pi}\sinh{k\pi \over 2\gamma}.
\end{equation}
In the same way, we could have computed the threshold velocity for
a given strength of the potential, but we prefer to check our predictions
this way because the presence of $\gamma$ makes the other possibility
more complicated. We compared this prediction to numerical simulations.
We show an example of this comparison in Fig.\ \ref{coll}, where we
study the propagation of a kink with initial velocity 0.2
(i.e., in a non-relativistic situation) on a potential
of a moderately long wavelength. The predicted threshold for kink
propagation as given by Eq.\ (\ref{thre}) is $\epsilon_{\scriptstyle\rm
thr}=0.0424$; from Fig.\ \ref{coll} we see that the numerical result is
bounded by $0.043 < \epsilon_{\scriptstyle\rm thr} < 0.0435$, meaning that
the error in our prediction is at most of 2\% which is quite satisfactory.
We have checked several other cases, which we summarize in
Fig.\ \ref{collsumm}; an excellent agreement is always found,
even for very large values of the perturbation potential.

To conclude this section on the collective coordinate treatment of the
problem, we discuss another prediction of it that it is numerically
verified, relativistic effects playing now the relevant part.
By looking again at Eq.\ (\ref{veff}), it can be realized that
the presence of $\gamma$ in the potential may give rise to singularities
in the neighborhood of the maximum velocity, $v=1$. To check whether this
is so, we integrated numerically the
ODE obtained from applying second Newton's law to $V_{\scriptstyle\rm
eff}$, and found that if the initial conditions were those of a particle
starting at the top of a large potential, the velocity of that particle
would grow as it slides down the potential; of course, if the potential
is large enough, the velocity can reach $v=1$: In those cases the
numerical integration of the ODE broke
down. The question then arises whether this is an artifact of our
collective coordinate approach or there is a related phenomenon in the
full PDE. We actually found that this ODE prediction is verified,
as we shown in Fig.\ \ref{top}. In this simulation, the initial condition
was a kink at a maximum of the potential
with initial velocity 0.1 so as to start the motion from the stable
point. As it moves to the nearest well, it accelerates and, eventually,
its velocity becomes very close to 1, implying that the kink cannot
accelerate further. Then it is trapped at an intermediate point of
the potential, instead of continuing its motion to the bottom. The
existence of this counterintuitive phenomenon shows in a very dramatic
way the value of a simple approach like the collective coordinate
formalism to help understand complicated nonlinear phenomena.

\section{Length scale competition.}

\subsection{Numerical experiments.}

The numerical and theoretical analysis described so far allowed us
to achieve a quite good understanding of the periodically perturbed
sG problem, at least of the basic phenomenology. With that background
in mind, we then turned to the initial motivation for this work, namely
to study kink propagation in periodic potential as a step towards
the much more complicated problem of breather propagation on periodic
media \cite{sG}. In principle, we did not
expect length scale competition to arise in this problem, as kinks
are very robust objects (which is further confirmed by our above
results, in particular by the success of the collective coordinate
approach). On the other hand, opposite to the case of the breather,
kink widths do not vary much when changing the only parameter
governing it, the kink velocity. Of course, when approaching the
maximum velocity, Lorentz contraction of the kinks will make them
vanishingly small, but that is a regime in which is very difficult
and time consuming to carry out good numerical simulations, so we
did not address the problem in that limit. Therefore, the kinks we
are usually dealing with have a width of about 6 in our dimensionless
units. Our purpose was to perform some simulations
in the high perturbation regime to see whether any light could be
shed on
the related breather problem.

The numerical experiments we made were as follows. We studied
several potentials of different wavelengths, ranging from 0.5 to 20,
i.e., from much smaller than the kink width to roughly three times
its width. The initial condition was always a kink at one top
of the potential; different velocities were considered. A summary of these
experiments in the most interesting range of potential wavelengths
is shown in Figs.\ \ref{competition} and \ref{shapes}. Figure
\ref{competition} shows the motion of the center of mass of the
kink on different potentials. The motion in the small or large
wavelength limits has been already discussed and it is again seen
here. However, a more interesting phenomenon is also evident,
namely kink trapping [or even reflection in the case of
wavelength 3, see Fig.\ \ref{shapes}(b)];
this trapping was not to be expected because kinks
start from a maximum of the potential and with a large initial velocity
(in Fig.\ \ref{competition}, it was 0.5). Regarding this point,
we have to stress here
that this trapping is of a different kind that the one discussed in
the preceding section, which was clearly a non-radiative process.
Besides, the trapping
depends crucially on the wavelength of the potential. Thus, for
instance, in the case of wavelength 2 [Fig.\ \ref{shapes}(a)],
the kink is able to propagate over six wells, whereas when the
wavelength was 4 or 5 [Figs.\ \ref{shapes}(c), (d)], it was trapped
already on the second well.
This dependence clearly indicates that length scale competition is
to same extent present also in the kink problem. This hint is
further supported by our previous work for the breather case \cite{sG},
which showed that this competition was most effective when the wavelength
of the perturbation was around half the breather width or slightly larger.
This is also the case in these simulations. Another common feature
between both problems is that the outcome of a simulation depends
very sensitively on the initial condition. This can be understood
from the reflection case in Fig.\ \ref{shapes}(b): For the kink to jump
back over one potential wavelength it is necessary that it meets the
radiation it left behind in the appropriate phase to gain energy from
it, and this evidently depends crucially on the initial velocity, as
we checked in our simulations. Hence, we conclude that these
phenomena are a novel manifestation of length scale competition.

\subsection{Linear stability analysis and discussion.}

The numerical findings we have described in the preceding subsection
are of great importance: The existence of processes governed by length
scale competition in the sG kink case opens the possibility of a
deeper study of the mechanisms through which this competition affects
the kink evolution. The relevance of this comes from the fact that,
when studying the sG breather problem \cite{sG}, we were not able
to carry out this deeper analysis due to the more complicated nature
of the breathers, namely their intrinsic internal dynamics which
severely complicate linear stability analysis.
But after showing that this competition also affects
kinks, we can certainly study their linear stability analysis, and,
consequently, obtain for the first time an understanding of the
mechanisms underlying length scale competition.

We tested the stability of the analytical continuum solutions as well as
of the numerical solutions in the following way. Let the solution to
discretization of the sG equation (\ref{PsG}) be $u_i = u^{(0)}_i +
v_i$, where $u^{(0)}_i$ is either the discretized version of the
continuum kink or the true minimum energy static solution of the
perturbed sG equation (\ref{PsG}) obtained numerically, and where
$v_i$ is a small discrete-valued function whose time dependence is
given by $\sin(\omega t)$; the index $i$ runs over the $N$ points of the
discrete lattice. The discrete version of the perturbed
problem (\ref{PsG}) is given by
\begin{equation}
\label{discretesG}
\ddot{u_i} - a^{-2} (u_{i+1} - 2u_{i} + u_{i-1}) +
[1+\epsilon\cos(kai)]\> u_i =0,
\end{equation}
where $a$ is the lattice spacing. Substituting the proposed form for
$u_i$ in Eq.\ (\ref{discretesG}) and linearizing we get
\begin{equation}
\label{ls}
\bbox{\Omega} {\bbox v} $ = $ \omega ^2 \bbox{v},
\end{equation}
where $\bbox{v}$ is the vector containing the $v_i,\>i=1,\ldots,N$ and
$\bbox{\Omega}$ is an $N\times N$ matrix given by
\begin{equation}
\bbox{\Omega}_{ij} = \left\{ \begin{array}{ll}
2+[1+\epsilon\cos(kai)]\cos\left(u^{(0)}_i\right), & \mbox{if }i= j,\\
-1, & \mbox{if }i=j\pm 1, \\ 0, & \mbox{otherwise.} \end{array}
\right.
\end{equation}
In this formulation, the modes for the linear excitations around
the shape $u^{(0)}$ are obtained simply by solving for the eigenvalues
$\omega^2$ of the matrix $\bbox{\Omega}$. We did this for all the
wavelengths we were studying, taking for $u^{(0)}$
the exact continuum sG kink at the top or at the bottom
of the potential, as well as the numerically obtained solutions
at similar positions.

The results for our numerical linear stability analysis
are shown in Fig.\ \ref{spectra}
for some of the relevant wavelengths \cite{referee}.
There are a number of interesting
features which deserve comment. First, let us consider
the spectra for the exact continuum shapes. When placed at the top of
the potential, this gives rise to a negative lowest eigenvalue
$\omega^2$, indicating that this continuum kink is not an exact
solution of the discrete problem and has a tendency to relax to the
correct one by emitting a burst of radiation.
The shape at the bottom of the potential does not show
this negative eigenvalue but, instead, a single bound state with
frequency $\omega^2\approx 0.3$. This corresponds to
a shape mode, similar to that present in unperturbed $\phi^4$
kinks, and it actually shows up in simulations: Initial data placed
at a potential well (an exact continuum kink) exhibit a static center
of mass but a general oscillating shape. This is easily understood
if one realizes that in this range of potential wavelengths, different
parts of the kink undergo the action of very different
perturbation values (that can be even positive or negative contributions).
In response to these gradients, the continuum kink oscillates.
These isolated states characteristic of the continuum kinks disappear
when we analyze static numerical solutions. It is seen from Fig.\
\ref{spectra} that in that situation spectra are composed of bands.
Actually, this is a common feature to all analyzed shapes, including
the continuum ones, and it could be expected in view of the following
argument: Far from the kink center, which only spans a small fraction of
the lattice sites, the nonlinear contribution to the linearized discrete
problem for $v_i$ vanishes, and one is left with what is a
standard Floquet (Bloch) problem. The corresponding structure is very
well known, and in fact it is very much like the ones we show here,
with gaps at positions that depend on the potential wavenumber and
gap amplitudes that depend on the potential strength ($\epsilon$).

By
comparing the spectral structure we have obtained to the outcome
of the numerical simulations (Fig.\ \ref{spectra}) and especially
to the radiation emission (Fig.\ \ref{shapes}), the mechanism for
kink destabilization can be inferred. In the case of small wavelength
potentials, the spectrum is very similar to the unperturbed sG one
(see Fig.\ \ref{spectra}a). Therefore, the behavior of the kink
is very similar to the continuum one moving in a discrete lattice,
the periodic potential then being nothing but the Peierls-Nabarro
barrier, as we already mentioned.
In that case the kink is known to radiate \cite{Michel}
and correspondingly decelerate until it is eventually trapped in
a potential well (Fig.\ \ref{shapes}b). When the wavelength is
smaller, the height of the effective potential seen by the kink is
so small [due to the $\sinh$ term in Eq.\ (\ref{veff})] that this
effect is hardly noticeable (hence the perfect constant motion of
the kink on the smallest wavelength potential in Fig.\
\ref{competition}). As the wavelength increases, more and more modes
move below the phonon band, inducing shape oscillations of the kink,
and as a consequence of this motion, long wavelength radiation is
emitted [clearly seen in Figs.\ \ref{shapes}(a) and (b)]. This is possible
because in those cases there is still a large number of available
modes just above the phonon band. Note that the lower limit of the
phonon band is given by
$\omega=1$; lower frequencies are localized, because they cannot
propagate in the system far from the kink, where the spectrum structure
is essentially the unperturbed one.
This combined effect induces a rapid
destabilization of the kink and its trapping. Finally, if the wavelength
is further increased, all the first band eventually moves below the
phonon band (Fig.\ \ref{spectra}d), and hence long-wavelength emission
is strongly suppressed (Fig.\ \ref{shapes}d), which stabilizes the
kink, making possible its propagation. The effect of the shape modes
coming from the first band is still revealed by the kink shape oscillations
[see Fig.\ \ref{shapes}(d)]. We believe that this interpreation
clearly explains the mechanisms leading to the appearance of length
scale competition in sG kinks. To seek further evidence,
we looked for the approximate value of the potential
wavelength at which the last mode in the first band crosses the phonon
band. It is shown in Fig.\ \ref{th} that this happens for a potential
wavelength between 5.2 and 5.3. Were our hypothesis true, kinks would
not be able to propagate on the former potential and they would be
able to do it in the latter one. The numerical simulations shown in Fig.\
\ref{numth} confirm that this is indeed the case. Interestingly,
the radiation is quite different in both cases and, furthermore, for
the trapped kink trapping occurs at the second potential well,
indicating that the small number of modes available to radiate prevents
a very rapid decay of the kink. We thus conclude that our interpretation
is indeed correct. As a matter of fact, as this feature of the spectrum will
also be present when considering breather propagation (recall that the
reason for the appearance of the gaps is the perturbating potential
acting on the wings of the excitation, and the nature of the center becomes
less relevant), our explanation of length scale competition should
also apply to breathers. The results in Ref.\ \cite{sG} are in perfect
agreement with what we have described in this section.

\section{Conclusions.}

In this paper, we have studied kink propagation in 1D sG systems
with a spatially periodic modulation of its characteristic frequency.
We have shown numerically that kinks can propagate steadily and mostly
undisturbed, even for large amplitudes of the perturbation. Disagreement
with analytical predictions previous to this work is resolved through
a new perturbative calculation. By this means, the radiative divergence
is shown
to be similar to an ``infrared'' divergence. Our calculation provides us
with a good physical understanding of the problem of free kink propagation
in the periodically modulated sG system, which had not been
obtained from the previous ones, of a more formal character. A comment
is in order here, regarding the fact that now the mathematical
foundations of perturbation theory for solitons are already established
(mostly by pioneering works like Refs.\ \onlinecite{raro,Boris1}),
the emphasis of that perturbation theory should be focused on their
physical implications. Therefore, perturbative
calculations in nonlinear equations should be regarded as speculative
if they are not verified through numerical simulations, and,
most importantly, if their physical meaning is not fully established.
That is one of the most important points of this work: After an
appropriate perturbative calculation, and by carefully
considering the dimensions of the problem, we have been able to identify
the underlying physical reason for that divergence as a
resonance with the lowest, infinite wavelength phonon mode. Moreover,
motivated by our perturbative results, we have developed a collective
coordinate approach to this problem that describes in a quantitatively
correct way the main features of kink propagation, showing that the
already known \cite{sG} ``bare'' and ``renormalized'' particle limits
apply also in this case. The collective coordinate equations turn
out to predict counterintuitive phenomena whose existence is confirmed
by numerical simulations, namely kink trapping at intermediate
points in the potential. This is an important success of the technique.
Finally, we have shown that length scale competition arises unexpectedly
in this problem, which afforded us the opportunity to increase our
understanding of this ubiquitous phenomenon \cite{sG,NLS}. By a detailed
linear stability analysis, we have identified the mechanism
by which length scale competition arises as coming from the band
structure induced by the perturbation potential. Again, the predictions
of our theory have been fully confirmed (quantitatively) by
the corresponding numerical simulations.

The global picture that emerges from this work is that, once again, sG
kinks behave basically like particles and a collective coordinate
approach can be more faithful than complicated perturbative results.
Length scale competition phenomena are only relevant in a large
amplitude
perturbation regime and therefore do not interfere with this picture
in most situations.
In this respect, it has to be noted that a perturbative calculation
describes everything beyond the center-of-mass dynamics:
Extended (background) contributions (see, e.g., the third reference in
\cite{viejos}); shape changes localized around the (moving) kink; and
the radiation, i.e., emission from the kink. It is crucial to separate
and identify these physically different effects if one is to properly
understand the dynamics of the considered system.
Our results are likely
to be general for kink-bearing models in view of the related results
of \cite{Madrid} on the spatially periodically perturbed $\phi^4$ model.
Some remarks are in order regarding this related problem. The same kind
of divergence is predicted by a pertubation theory similar to the one
used here (see \cite{yo2} for details on this approach), and again,
numerical simulations show
evidence of the unphysical character of the divergence: It can be
seen from Fig.\ 4 of \cite{Madrid} that as the velocity of a decelerating
kink goes through the threshold nothing special occurs. As a matter of
fact, most of the reasoning we have used in the present study applies
also to that work (with an additional feature coming from the shape
mode of the unperturbed $\phi^4$ kink),
thus reinforcing the generality of our results.
On the other hand, the discovery of length scale competition
for sG kinks has allowed us to understand the
underlying physical reasons,
and helped us to gain insight on related results for the sG
breather \cite{sG}. We believe that the
mechanism we have identified in this work is of a very general character,
and it will be important to have more work on related systems
to check our predictions. As a final remark, we want to stress that we
have provided a quite thorough description of the features of free
kink propagation in a periodically modulated sine-Gordon system, and
that we have been able to provide a consistent, physical framework
to understand this ``canonical'' problem, that will be of help
in dealing with more
complex situations.

\section*{Acknowledgments}

We are indebted to Rainer Scharf, David Cai, and Maxi San Miguel
for enlightening conversations on this research.
A.S.\ is supported by a Ministerio de Educaci\'on y
Ciencia (Spain)/Fulbright postdoctoral
scholarship, by Direcci\'on General de Investigaci\'on
Cient\'\i ifica y T\'ecnica (Spain) through project PB92-0378,
and by the European Community ({\em Network}\/ on Nonlinear
Spatio-Temporal Structures in Semiconductor, Fluids, and
Oscillator Ensembles).
He also thanks the Los Alamos National Laboratory for
warm hospitality and a productive atmosphere.
Work at Los Alamos is performed under the auspices of
the U.S.\ Deparment of Energy.

\begin{figure}
\caption[]{Absence of radiative divergence
for kinks propagating in the spatially
periodic sG model. Parameters are: (a)
$k=2\pi/5$, initial velocity $v=v_{\scriptstyle\rm
thr}=0.387726\ldots$; (b), $k=\pi$, initial velocity $v=v_{\scriptstyle\rm
thr}=0.091999\ldots$; (c), $k=2\pi$, initial velocity $v=v_{\scriptstyle\rm
thr}=0.024704\ldots$ [corresponding wavelengths are (a) 1, (b) 2,
(c) 5]. In all three cases,
$\epsilon=0.4$. The amplitude of the emitted
radiation is very small; due to the periodic boundary conditions, it can be
seen reentering the simulation interval without any appreciable
interaction
with the kink. Only half of the simulation interval is shown in
plots (b) and (c) to enlarge details. Time increases upwards with
final time $t=100$. The potential is indicated by the dashed line
(amplitude not to scale).}
\label{nodiv}
\end{figure}

\begin{figure}
\caption[]{Instantaneous center of mass positions
as obtained from the simulations
in Fig.\ \ref{nodiv}. Dot-dashed line: $k=2\pi/5$. Dashed line:
$k=\pi$. Solid line: $k=2\pi$.}
\label{centers}
\end{figure}

\begin{figure}
\caption[]{The pole structure of the radiation contribution.
Filled circles mark the location of the poles which give rise to
corrections localized around the soliton. Empty circles denote
the locations of pole $z_1$ as the velocity changes. Only when
$z_1$ becomes real ($v>v_{\scriptstyle\rm thr}$) does it originate
propagating wavelike corrections.
See text for further explanation.}
\label{poles}
\end{figure}

\begin{figure}
\caption[]{An
example of the verification of the collective coordinate approach
predictions. For a kink starting from a mid-point of a potential
of wavelength 20, with velocity 0.2, the predicted threshold for
propagation is $\epsilon_{\scriptstyle\rm thr}=0.0424$. (a) $\epsilon =
0.43$ and the kink propagates; (b) $\epsilon=0.435$ and the kink
is reflected by the potential maximum;
(c) center of mass motion for better comparison
of both cases; solid line corresponds to the simulation in (a) and
dashed line to that in (b). Final time is $t=200$. Notice the absolute
absence of radiation in this phenomenon.}
\label{coll}
\end{figure}

\begin{figure}
\caption[]{Comparison of the effective potential prediction with
numerical simulations
for a number of cases. Solid line: Threshold for initial velocity 0.2
as a function of the potential wavenumber. Dashed line: The same for initial
velocity 0.5. Diamonds and crosses are points obtained from the numerical
study.
Error bars in the numerical threshold are of the order of the size of
the symbols.}
\label{collsumm}
\end{figure}

\begin{figure}
\caption[]{Trapping of a kink at an intermediate position in
the potential instead
of at the bottom. $\epsilon=2$ and the kink starts with velocity 0.1.
The amount of emitted radiation is large due to the fast acceleration
of the kink in this large perturbation. (a) time evolution of the
kink; (b) time evolution of the center of mass. The neighboring minimum is
at $x=-2.0$ (indicated by the dashed line),
and the kink oscillates around approximately $x=-2.5$.
Final time is $t=50$.}
\label{top}
\end{figure}

\begin{figure}
\caption[]{Length scale competition. Center of mass evolution of a kink
on potentials of different wavelengths, always with $\epsilon=0.7$ and
initial velocity 0.5. Wavelengths are: $\Diamond$, 1; $+$, 2; $\Box$, 3;
$\times$, 4; $\bigtriangleup$, 5; and $*$, 6. See text for explanation.}
\label{competition}
\end{figure}

\begin{figure}
\caption[]{Time evolution for some of the kinks in Fig.\
\protect\ref{competition}.
Wavelengths are (a) 2; (b) 3; (c) 4; and (d) 6. Notice the different
kind of radiation emitted in each case. Notice also that the kink in
(b) goes back over one maximum of the potential probably due to resonant
interaction with radiation left behind. Final time is $t=50$.}
\label{shapes}
\end{figure}

\begin{figure}
\caption[]{Frequency spectra for kinks on periodic potentials. See
text for details on how is it computed. Spectra for numerical shapes
are plotted with $\Diamond$ (potential maximum), and $\Box$
(potential minimum). Spectra for analytical shapes are plotted
with $+$ (potential maximum) and $\times$ (potential minimum).
Wavelengths are (a) 2; (b) 3; (c) 4; and (d) 6. Notice in this
last case that all the first branch of spectrum is below the phonon
band.}
\label{spectra}
\end{figure}

\begin{figure}
\caption[]{Spectra for kink on
periodic potentials of wavelength 5.2 ($\Diamond$) and 5.3 ($\times$).
Between this two values all the first band of spectrum occurs below
the phonon band.}
\label{th}
\end{figure}

\begin{figure}
\caption[]{Numerical
verification of the threshold for length scale competition.
Evolution of a kink on a potential of wavelength (a) 5.2, or (b)
5.3. (c) shows the motion of the center of mass in both cases for
better comparison;
solid line corresponds to the simulation in (a) and
dashed line to that in (b)}
\label{numth}
\end{figure}


\begin{references}
\bibitem{referee} It is possible to analyze the long wavelength limit of
the spectrum analytically, by means of an averaging technique, which leads
to the result thta the minimum frequency is given by $\omega_{min}^2=
1-(\epsilon^2/2k^2)$, in good agreement with our numerical calculation
of the spectrum as can be seen from the figures. We are indebted to the
referee who pointed out this result to us.
\end{references}
\end{document}